\documentclass[twocolumn,aps,pra,superscriptaddress]{revtex4-1}
\usepackage{amsfonts}
\usepackage{amsmath,amssymb}
\usepackage{mathtools}
\usepackage{extarrows}
\usepackage{bm}
\usepackage{graphicx}
\usepackage{overpic}
\usepackage{color}
\usepackage{mdframed}
\allowdisplaybreaks[1]

\makeatletter

\begin{document}

\title{Theory of Electron Spin Resonance Spectroscopy in Scanning Tunneling Microscopy}
\author{Lyuzhou~Ye}
\affiliation{Hefei National Research Center for Physical Sciences at the Microscale \& Synergetic Innovation Center of Quantum Information and Quantum Physics, University of Science and Technology of China, Hefei, Anhui 230026, P. R. China}
\author{Xiao~Zheng} \email{xzheng@fudan.edu.cn}
\affiliation{Department of Chemistry, Fudan University, Shanghai 200438, P. R. China}
\author{Xin~Xu} \email{xxchem@fudan.edu.cn}
\affiliation{Department of Chemistry, Fudan University, Shanghai 200438, P. R. China}
\affiliation{Hefei National Laboratory, Hefei, Anhui 230088, P. R. China}

\date{Submitted on February~7, 2024; resubmitted on August~1, 2024}

\begin{abstract}

The integration of scanning tunneling microscopy (STM) and electron spin resonance (ESR) spectroscopy has emerged as a powerful and innovative tool for discerning spin excitations and spin-spin interactions within atoms and molecules adsorbed on surfaces. However, the origin of the STM-ESR signal and the underlying mechanisms that govern the essential features of the measured spectra have remained elusive, thereby significantly impeding the future development of the STM-ESR approach. Here, we construct a model to carry out precise numerical simulations of STM-ESR spectra for a single hydrogenated Ti adatom and a hydrogenated Ti dimer, achieving excellent agreement with experimental observations. We further develop an analytic theory that elucidates the fundamental origin of the signal as well as the essential features in the measured spectra. These new theoretical developments establish a solid foundation for the on-demand detection and manipulation of atomic-scale spin states, with promising implications for cutting-edge applications in spin sensing, quantum information, and quantum computing.

\end{abstract}

\maketitle

{\it Introduction.} The precise measurement and control of spin states of single atoms and molecules \cite{Rau2014May,moreno-pineda_measuring_2021,Heinrich2021Dec}
serve as the foundation for cutting-edge applications in spintronics \cite{bogani2008molecular,coronado2020molecular,moreno2021measuring}, quantum sensing \cite{Choi2017May,Zhang2022Jan}, quantum information \cite{natterer_reading_2017,singha_engineering_2021}, 
and quantum computing \cite{chen_harnessing_2022,Bi2023Feb}.
Over the last decade, the synergistic integration of electron spin resonance (ESR) technique with scanning tunneling microscopy (STM) has fundamentally transformed the ability to detect and analyze spin states in atoms and molecules adsorbed on surfaces, offering unprecedented spatial and energy resolution
\cite{loth_measurement_2010,baumann_electron_2015,Choi2017May,natterer_reading_2017,
Paul2017Apr,PhysRevLett.119.227206,Bae2018Nov,Willke2018Feb,
Willke2019Nov,PhysRevLett.122.227203,Seifert2020Oct,Yang2021Feb,Willke2021Nov,singha_engineering_2021,Zhang2022Jan,chen_harnessing_2022,Bi2023Feb,Kot2023Oct,yang_coherent_2019,veldman_free_2021,Ast2024Mar}.
The STM-ESR spectroscopy provides profound insights into spin-related properties of single atoms and molecules,
including spin-state transitions and spin couplings between magnetic centers. 
Recent strides made in STM-ESR have enabled coherent control over the temporal evolution of
atomic spin states \cite{yang_coherent_2019,veldman_free_2021},
leading to the establishment of an atomic-scale multi-qubit platform \cite{wang_atomic-scale_2023}, further advancing the frontier of quantum technologies.

\begin{figure}[htbp]
\centering
\includegraphics[width=1.\columnwidth]{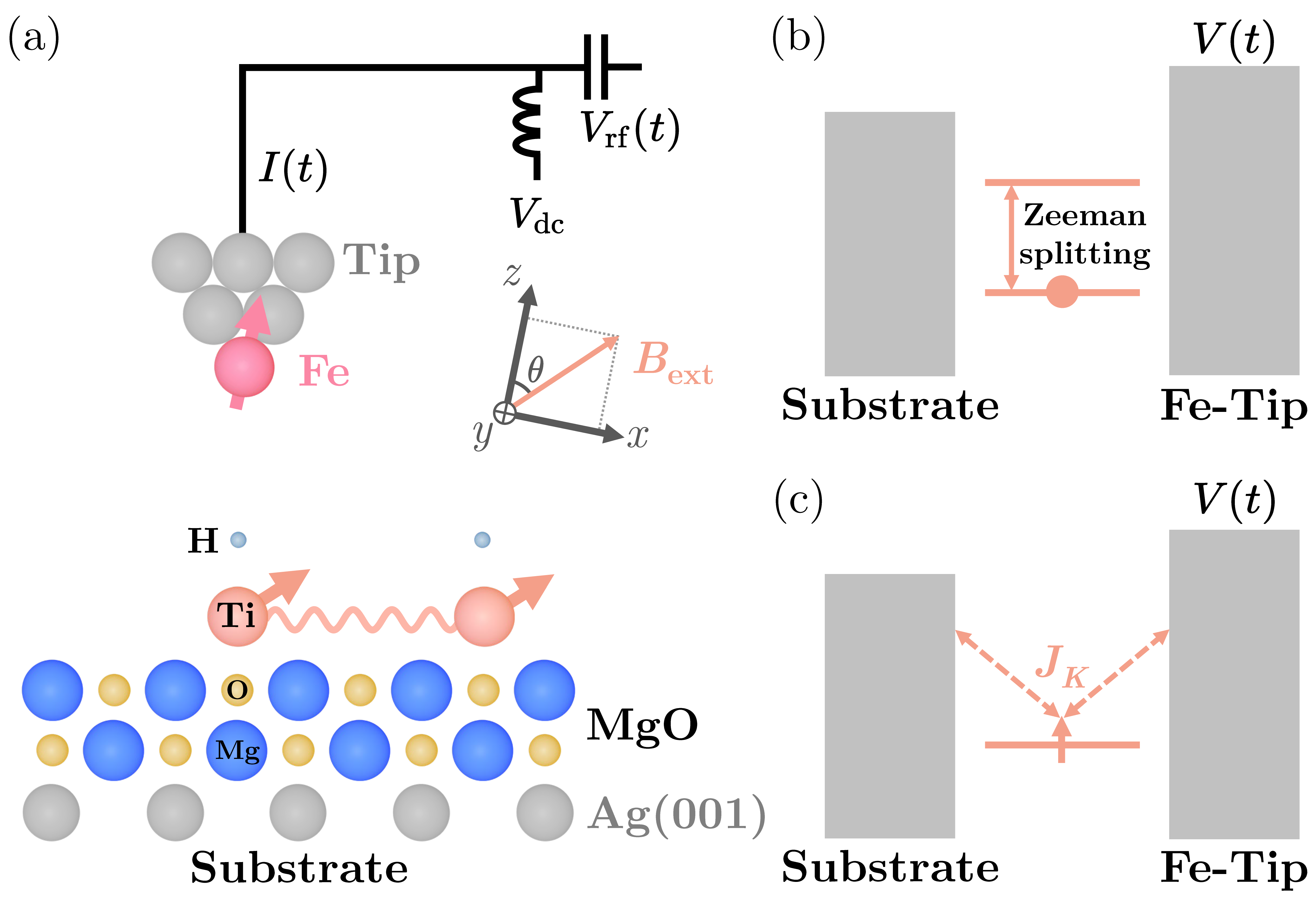}
\caption{(a) Sketch of the experimental setup,
where the STM tip with Fe atoms at its apex is utilized to probe spin-$1/2$ hydrogenated Ti adatoms.
The substrate consists of a bilayer MgO grown on Ag(001).
The spin moment of the Fe atom is along $\hat{\boldsymbol z}$. An external magnetic field ${\boldsymbol B}_{\rm ext}$ is applied in the $xz$-plane, 
with angle $\theta$ determining its direction relative to $\hat{\boldsymbol z}$.
A bias voltage, composed of $V_{\rm dc} $ and $V_{\rm rf}(t)$, is applied to the tip,
resulting in the time-varying electric current $I(t)$. 
(b) Illustration of the AIM, and (c) illustration of the Kondo model, where $J_K$ denotes the Kondo exchange interaction.} 
\label{fig1}
\end{figure}

Despite the remarkable success of the STM-ESR technique, there is a persistent ambiguity regarding the origin of the STM-ESR signal \cite{Delgado2017Feb,delgado_theoretical_2021}.
Unlike conventional ESR experiments that employs alternating magnetic fields to probe macroscopic samples \cite{weil2007electron}, the STM-ESR technique makes use of  an alternating current (ac) voltage as the driving source. 
This raises two fundamental questions: 
(1) What is the nature of the quantum dynamics that leads to the generation of STM-ESR signal?
(2) How is the underlying quantum dynamics imprinted in the electric current, ultimately yielding the characteristic spin resonance signature?

To address these questions, various mechanisms 
\cite{baumann_electron_2015,ferron_single_2019,PhysRevB.96.205420,Seifert2020Oct,reina_galvez_cotunneling_2019,PhysRevB.104.245435,PhysRevB.107.235404,PhysRevB.99.054434,PhysRevB.92.220418,Ast2024Mar,delgado_theoretical_2021} have been proposed to explore the dynamics of the nuclei, 
electric charges, and/or atomic spins driven by the ac voltage. 
The piezoelectric response mechanism \cite{baumann_electron_2015, PhysRevB.96.205420,ferron_single_2019,Seifert2020Oct} focuses on voltage-induced nuclear motions of surface samples, but it was suggested that these effects are relatively small and hence not a mandatory ingredient \cite{reina_galvez_cotunneling_2019,delgado_theoretical_2021}. 
The tunneling barrier modulation mechanism \cite{reina_galvez_cotunneling_2019,PhysRevB.104.245435,PhysRevB.107.235404} 
is based on the idea that the electron hopping integrals vary with time due to the ac voltage, but it predicts zero signal in specific ranges of the applied direct current (dc) voltage \cite{PhysRevB.107.235404}, which was not observed experimentally. 
The spin-transfer-torque mechanism \cite{PhysRevB.99.054434} highlights the dynamics of local spin moments driven by environmental fluctuations, but the predicted spectra exhibit a linewidth independent of the ac voltage magnitude, which is inconsistent with experimental observations \cite{Bae2018Nov,PhysRevLett.122.227203,Seifert2020Oct}.
Therefore, the existing theoretical understandings do not fully account for the key experimental features observed.

This Letter aims to deliver a definitive resolution to the outstanding questions. To this end, we construct a microscopic model that takes a simple form yet enables precise numerical simulations of the STM-ESR spectra. More importantly, we exploit the notion that the spin and charge dynamics are essentially decoupled, due to the pronounced separation between the charge and spin excitation energy scales on the sample
\cite{PhysRevLett.119.227206,yang_coherent_2019,Willke2021Nov}. This allows us to establish an analytic theory that elucidates the origin of the STM-ESR signal and comprehensively explicates the characteristic features observed in experiments.

{\it Microscopic modeling of the experimental setup.} 
Figure~\ref{fig1}(a) illustrates the STM-ESR experimental setup. The ESR signal is attained as the dc component of electric current driven by a time-dependent bias voltage $V(t) = V_{\rm dc} + V_{\rm ac}(t)$ \cite{PhysRevLett.119.227206,yang_coherent_2019,veldman_free_2021},
where $V_{\rm dc}$ and $V_{\rm ac}(t) = V_{\rm rf}\sin(\omega_{\rm rf} t)$ denote a constant dc voltage and a continuous-wave radio-frequency (rf) voltage, respectively. 

We adopt the Anderson impurity model (AIM) to represent the STM junction.
The total Hamiltonian consists of three parts, i.e., $H_{\rm T} = H_{\rm imp} + H_{\rm res} + H_{\rm coup}$. 
Here, the impurity Hamiltonian $H_{\rm imp}$ describes the magnetic orbitals of the surface sample. 
For a hydrogenated Ti atom on the MgO/Ag surface, first-principles studies \cite{ferron_single_2019,steinbrecher_quantifying_2021,PhysRevB.104.174408} have shown 
that the ${\rm d}_{x^2-y^2}$ orbital on the Ti adatom is singly occupied and contributes predominantly to the local spin moment, which can be represented by
\begin{equation}
H_{\rm imp} = \epsilon \left( \hat{n}_{\uparrow} + \hat{n}_{\downarrow} \right)  
+ U \hat{n}_{\uparrow} \hat{n}_{\downarrow}
+ H_{\rm Zeeman}, \label{AIM}
\end{equation}
where $\hat{n}_{\sigma}$ is the spin-$\sigma$ electron number operator ($\sigma = \uparrow$ or $\downarrow$),
and $\epsilon$ and $U$ are the orbital energy and the electron-electron interaction energy, respectively. 
$H_{\rm Zeeman} = g \mu_B  {\boldsymbol B}_{\rm ext} \cdot \hat{\boldsymbol S}$, where $g$ is the electron Zeeman factor, $\mu_B$ is the Bohr magneton, $\hat{\boldsymbol S}$ is the local spin operator,
and ${\boldsymbol B}_{\rm ext}$ is the external magnetic field that includes the field generated by the permanent spin moment of the Fe atom at the tip apex.

In the AIM, $H_{\rm res}$ represents the tip and substrate that are taken as non-interacting electron reservoirs (res), while $H_{\rm coup}$ describes the impurity-reservoir couplings (coup). Since the reservoirs follow Gaussian statistics, their impact on the impurity is fully captured by the hybridization functions, $\Gamma_{\alpha\sigma}(\omega) = \Gamma_{\alpha\sigma} D_\alpha(\omega)$, where $\Gamma_{\alpha\sigma}$ and $D_\alpha(\omega)$  represent the hybridization strength and dimensionless density of states of $\alpha$-reservoir ($\alpha=t,s$), respectively. In the following, $t$ and $s$ will be used as the subscripts to represent the tip and substrate, respectively. 
The tip spin polarization is oriented in the $\hat{\boldsymbol z}$ direction, and its extent is quantified by $\chi$, defined as $\chi \equiv \left( \Gamma_{t\uparrow} - \Gamma_{t \downarrow} \right) / \Gamma_{t}$, with $\Gamma_t= \Gamma_{t\uparrow} + \Gamma_{t\downarrow}$.

Revealing the role of the applied voltage in generating the spin resonance signatures is crucial for unraveling the underlying mechanisms of STM-ESR spectroscopy.
As shown in Fig.~\ref{fig1}(b), the voltage modulates the electrochemical potential of the tip, which drives the dynamics of the charge and spin states associated with the magnetic orbital of the surface sample. 
Alternatively, the influence of $V(t)$ on the quantum dynamics of the sample can be captured by a time-dependent phase factor in the electron hopping integrals \cite{PhysRevB.48.8487,jauho_time-dependent_1994, reina_galvez_cotunneling_2019,PhysRevB.104.245435,PhysRevB.107.235404}.

The parameters adopted in our microscopic model are determined based on experimental data \cite{PhysRevLett.119.227206,veldman_free_2021} and first-principles calculations \cite{ferron_single_2019,steinbrecher_quantifying_2021,PhysRevB.104.174408}.
The energy parameters pertinent to the charge dynamics, such as $U$, $\epsilon$, $V_{\rm dc}$ and $V_{\rm rf}$, are on the order of $0.01\sim 1\,{\rm eV}$. In contrast, those directly associated with the spin dynamics, such as the Zeeman splitting, exchange and dipolar spin couplings, impurity-reservoir hybridization $\Gamma_{\alpha \sigma}$, and temperature $T$, are on the order of $10^{-4}\sim 0.1\,{\rm meV}$.
By utilizing the Schrieffer-Wolff transformation \cite{PhysRev.149.491,kaminski_universality_2000}, these two disparate energy scales can be formally separated, leading to a spin-polarized Kondo model that focuses specifically on the spin degrees of freedom, as illustrated in Fig.~\ref{fig1}(c).

{\it Numerical simulation of STM-ESR spectra.} 
We perform precise simulations on the constructed AIM by employing the numerically exact hierarchical equations of motion (HEOM) method implemented in the HEOM-QUICK2 program \cite{Tanimura1989Jan, Jin2008Jun, PhysRevLett.109.266403, Ye2016Nov, Zha23014106, Zhang2024Jan}. 
The STM-ESR spectra are acquired by directly calculating and averaging the time-dependent electric current $I(t)$ in response to the applied ac voltage at different frequencies $\omega_{\rm rf}$.

The substantial disparity in energy scales between charge and spin dynamics presents a significant hurdle for the numerical simulations. Accurate propagation of the HEOM necessitates a very small time step, leading to computationally intensive simulations, particularly when extracting the dc component of the electric current. 
To tackle this difficulty, we scale down the parameters pertaining to charge excitations, such as $\epsilon$, $V_{\rm dc}$ and $V_{\rm rf}$, by the same factor, while ensuring that the pronounced distinction between the charge and spin excitation energy scales is preserved. This consistent scaling does not alter the lineshape of the resulting spectrum. Details are given in the Supplemental Material (SM) \cite{SupMat1}.

\begin{figure}[t]
\centering
\includegraphics[width=1\columnwidth]{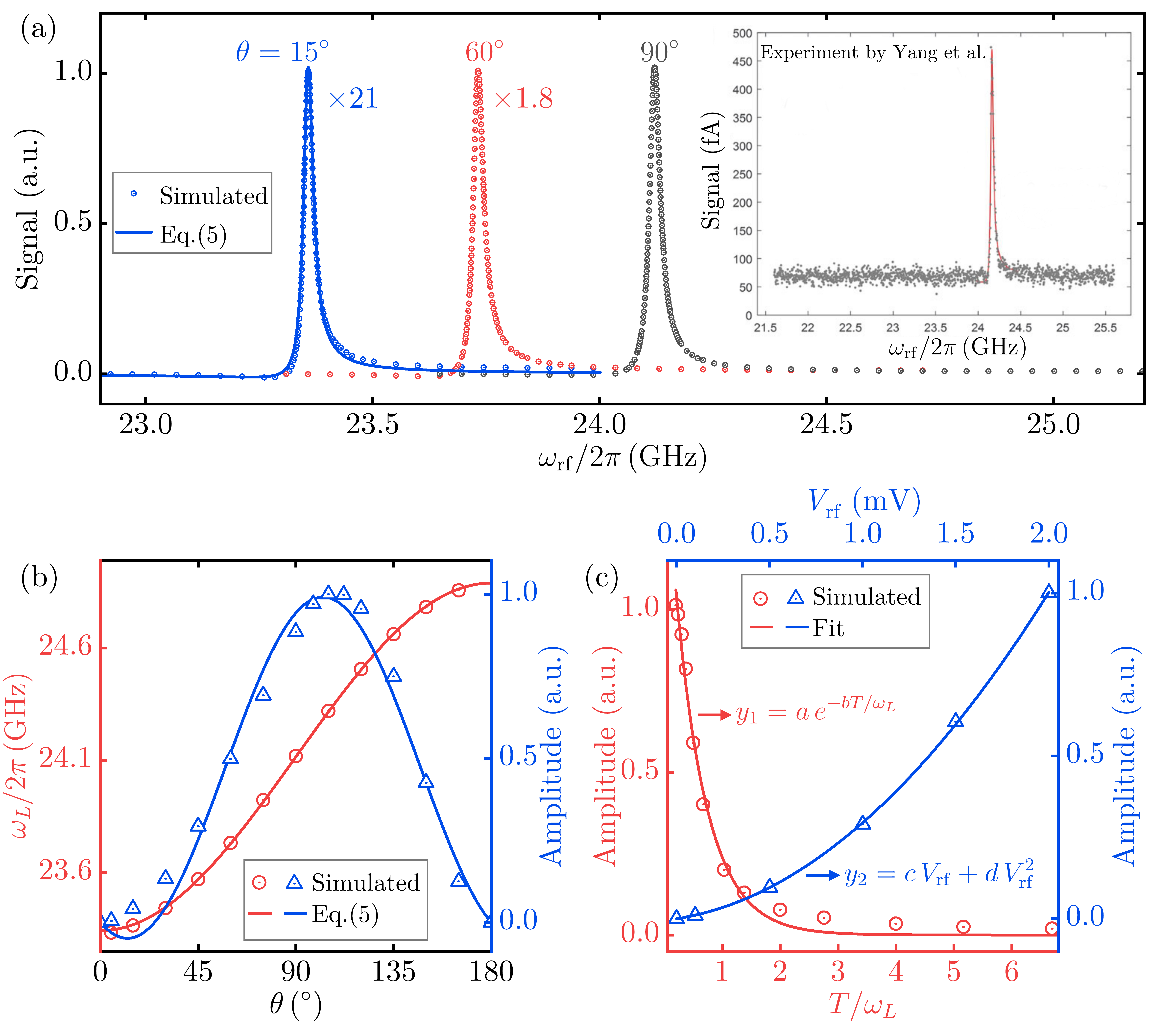}
\caption{(a) Simulated STM-ESR spectra (circles) of a single hydrogenated Ti adatom at different angles $\theta$ [see Fig.~\ref{fig1}(a) for definition of $\theta$],
with the analytic expression in Eq.~(\ref{stm-esr}) at $\theta = 15\,^\circ$ (line) shown for comparison.
All peaks are normalized to the same height.
The inset displays the experimental result obtained by Yang et~al. \cite{PhysRevLett.119.227206}.
(b) Variation of the resonance frequency $\omega_L$ (left axis) and amplitude (right axis) with respect to $\theta$,
where the lines represent the predictions by the analytic theory using Eq.~(\ref{stm-esr}). 
(c) Variation of the signal amplitude with respect to temperature $T$ (left and bottom axes) and $V_{\rm rf}$ (right and top axes) at $\theta = 90\,^\circ$. 
The data are fitted to an exponentially decaying function $y_1(\frac{T}{\omega_L})$ and nonlinear function $y_2(V_{\rm rf})$, respectively.
Other parameters adopted are: 
$\left|{\boldsymbol B}_{\rm ext}\right| = 0.86\,{\rm T}$, $\epsilon= -6\,{\rm meV}$,
$\Gamma_{s} = 0.2\,{\rm meV}$, $\Gamma_{t} =  0.01 \,{\rm meV}$, $\chi=0.9$ and $V_{\rm dc} = 0.4\,{\rm mV}$. 
Specifically, $T=1.2\,{\rm K}$ and $V_{\rm rf} = 0.2\,{\rm mV}$ are used in (a) and (b).} \label{fig2}
\end{figure}
%

Figure~\ref{fig2}(a) displays the simulated STM-ESR spectra of a single hydrogenated Ti adatom, featuring a resonance peak at the frequency $\omega_L$, which corresponds to the Zeeman splitting. The peak exhibits an asymmetric lineshape that resembles the experimentally measured spectrum shown in the inset. 
The simulated results reproduce several important features observed in experiments:
(i) The peak position and amplitude vary prominently with  angle $\theta$ [see Fig.~\ref{fig1}(a) for the definition of $\theta$], as depicted in Fig.~\ref{fig2}(b) \cite{PhysRevB.104.174408}.
Notably, the peak vanishes when $\bm B_{\rm ext}$ aligns parallel to the tip spin polarization at $\theta = 0$ or $180\,^\circ$. 
(ii) As demonstrated in Fig.~\ref{fig2}(c), the signal amplitude undergoes an exponential decay with increasing temperature $T$ \cite{Hwang2022Sep}, and exhibits a nonlinear dependence on $V_{\rm rf}$ in the low-bias regime \cite{Willke2018Feb}.

\begin{figure}[t]
\centering
\includegraphics[width=1.\columnwidth]{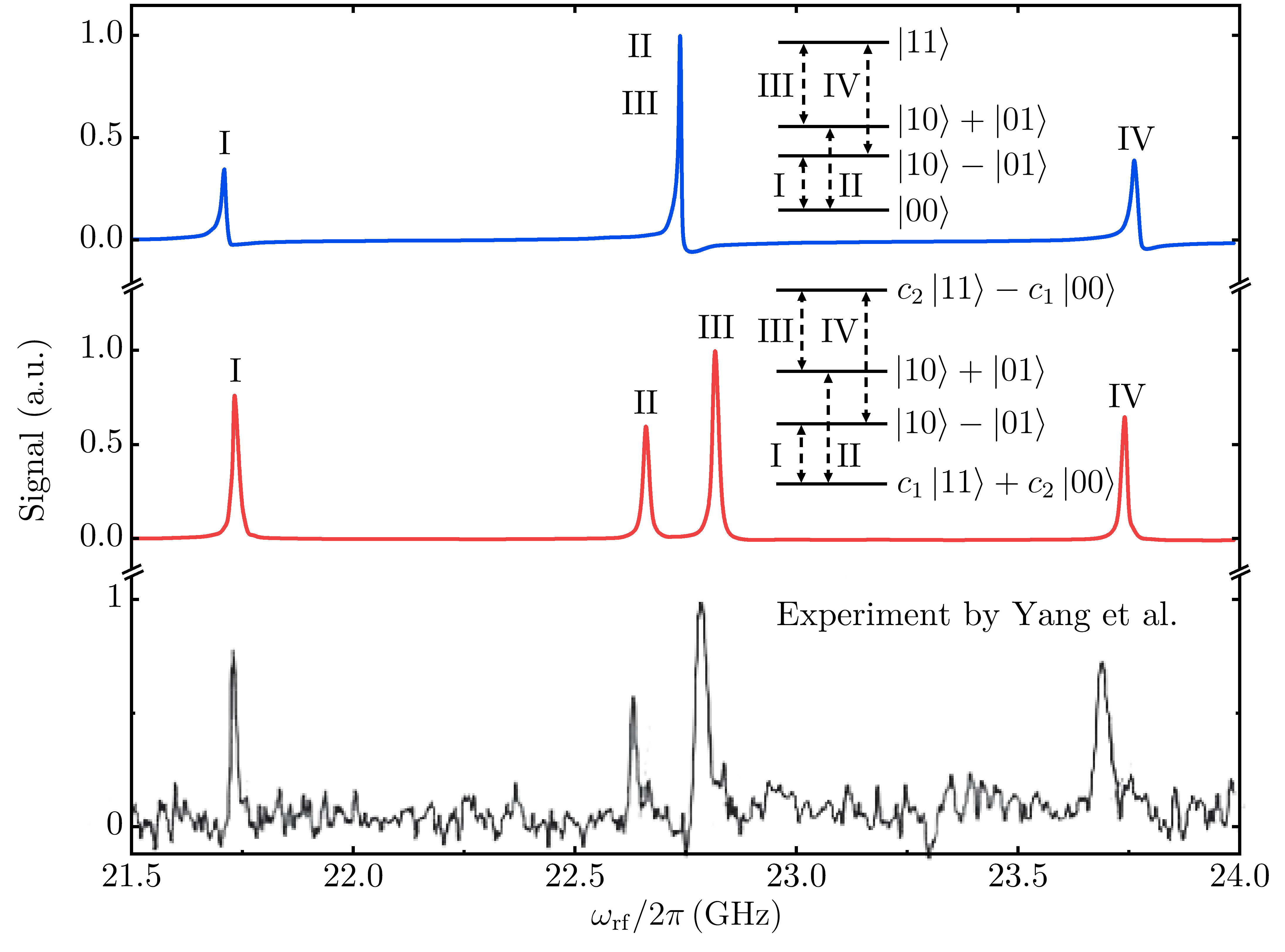}
\caption{Simulated STM-ESR spectra of a hydrogenated Ti dimer with $D = 0$ (upper panel) and $D = 0.105\,{\rm GHz}$ (middle panel), where $D$ indicates the magnitude of dipolar coupling.
The insets depict the lowest spin-states of the hydrogenated Ti dimer, where $0$ ($1$) indicates that the local spin aligns parallel (antiparallel) to ${\boldsymbol{B}_{\rm ext}}$.
The lower panel presents experimental data extracted from Ref.~\cite{PhysRevLett.119.227206}.
Other parameters adopted are $\epsilon= -6\,{\rm meV}$,
$\Gamma_{s1} =\Gamma_{s2}= 0.04\,{\rm meV}$,
$\Gamma_{t1\uparrow} = 9.5\times10^{-3}\,{\rm meV}$, $\Gamma_{t1\downarrow} = 5\times10^{-4}\,{\rm meV}$, $\Gamma_{t2\uparrow} = \Gamma_{t2\downarrow}= 0$, $V_{\rm dc} = 0.4\,{\rm mV}$, $V_{\rm rf}= 0.2\,{\rm mV}$, $\theta = 90\,^\circ$, $\left|{\boldsymbol B}_{\rm ext}\right| = 0.81\,{\rm T}$, $J =1.03\,{\rm GHz}$, and $T=1.2\,{\rm K}$. } \label{fig3}
\end{figure}

To demonstrate the unique capability of STM-ESR spectroscopy in probing extremely weak spin-spin couplings ($< 1\,{\rm GHz}$), we simulate the spectra for a hydrogenated Ti dimer. 
In experimental settings, the magnitudes of the exchange ($J$) and dipolar ($D$) couplings between the two spins depend sensitively on their relative distance and orientation, which can be precisely manipulated by using the tip as a mechanical tweezer \cite{PhysRevLett.119.227206,Bae2018Nov,Zhang2022Jan}.
In our simulation, the Fe-tip exclusively couples to the local spin at one hydrogenated Ti adatom. The couplings between the two local spins are described by the Hamiltonian,
$H_{\text{spin-spin}} = J \hat{{\boldsymbol S}}_1 \cdot \hat{{\boldsymbol S}}_2 + D ( 3 \hat{S}_1^{z} \hat{S}_2^{z} - \hat{{\boldsymbol S}}_1 \cdot \hat{{\boldsymbol S}}_2 )$, which constitutes an additional contribution to $H_{\rm imp}$.

Figure~\ref{fig3} depicts the simulated STM-ESR spectra alongside the experimental result \cite{PhysRevLett.119.227206}.
The resonance peaks are attributed to transitions between the lowest spin-states. In the insets, we observe that a nonzero $D$ alters the eigenstates of $H_{\rm imp}$, causing the central peak (upper panel) to split into two (middle panel).
Remarkably, the simulated spectrum obtained with a finite $D$ exhibits excellent agreement with the experimental result (lower panel) \cite{PhysRevLett.119.227206}.

The remarkable consistency between the numerical simulations and experimental observations, as evidenced in Figs.~\ref{fig2} and \ref{fig3}, unequivocally validates our constructed microscopic model and the role of the ac voltage in modulating the tip's electrochemical potential. To further elucidate the origin of the spin resonance signal and comprehend the significant characteristics depicted in Fig.~\ref{fig2}, we now introduce an analytic theory.

{\it Analytic STM-ESR theory.} 
We focus on the scenario where the charge and spin excitation energy scales of the measured sample are distinctly separated, allowing the charge and spin dynamics to be readily decoupled.
To demonstrate this, we consider a sample with a singly occupied magnetic orbital. In the large-$U$ limit, the Schrieffer-Wolff transformation \cite{PhysRev.149.491,kaminski_universality_2000} converts the AIM into a Kondo model, which describes a local spin moment of $S=1/2$ coupled to the reservoirs \cite{SupMat1}. The local spin is represented by the Hamiltonian $H_{\rm spin} = H_{\rm Zeeman}$, and its spin-exchange couplings with the reservoirs are given by
$H_{\rm int}= H_{ss} + H_{tt} + H_{st} + H_{ts}$, where
\begin{equation}
H_{\alpha \alpha'} =  \sum_{ikk'\sigma\sigma'}
J_{\alpha \alpha'kk'\sigma\sigma'}  \hat{s}^{i}_{\alpha \alpha'kk'\sigma\sigma'} \hat{S}^{i\dagger}.  \label{Hint}
\end{equation}
Here, $i=\left\{+,-,z\right\}$, and $J_{\alpha \alpha'kk'\sigma\sigma'}$ is the Kondo exchange coupling strength. $\hat{s}^{i}_{\alpha \alpha' kk'\sigma\sigma'} \equiv 
\hat{d}^\dagger_{\alpha k \sigma} \tau^i_{\sigma\sigma'} \hat{d}_{\alpha'k'\sigma'}$, 
where $\hat{d}^\dagger_{\alpha k \sigma}$ ($\hat{d}_{\alpha k\sigma}$) is the creation (annihilation) operator of $\alpha$-reservoir and $\tau^i$ is the Pauli matrix. 
$H_{\alpha \alpha'}$ represents the transition between reservoir states $\left|\alpha k\sigma\right\rangle $ and $\left|\alpha'k'\sigma'\right\rangle$ mediated by the magnetic orbital. 
Specifically, $H_{ss}$ and $H_{tt}$ result in Kondo screening of the local spin by the conduction electrons in the reservoirs, while $H_{st}$ and $H_{ts}$ represent electron transport between the substrate and tip through electron co-tunneling events. 
It is this electron transport that gives rise to an electric current, which encompasses the STM-ESR signal.
Similar co-tunneling Hamiltonian has been given in Ref.~\cite{reina_galvez_cotunneling_2019}.

The reservoirs influence the spin dynamics of the sample through both a mean-field effect and fluctuations. The mean-field effect originates exclusively from the spin-polarized Fe-tip, expressed as $\langle H_{\rm int}(t) \rangle_{\rm res}= \langle H_{tt}(t) \rangle_{\rm res} = g \mu_B B_{\rm tip}^{\rm eff}(t) \hat{S}^z$, while the fluctuations are represented by $H'_{\rm int}(t) = H_{\rm int}(t) - \langle H_{\rm int}(t) \rangle_{\rm res}$.  
Here, $\langle \hat{O} \rangle_{\rm res} \equiv {\rm tr}_{\rm res} (\hat{O} \rho^{\rm eq}_{\rm res} ) $ 
denotes the statistical average of the operator $\hat{O}$ for a reference equilibrium state, with $\rho^{\rm eq}_{\rm res}$ being the equilibrium density matrix of the decoupled reservoirs. 
Note that the spin-transfer-torque mechanism \cite{PhysRevB.99.054434} only considers the fluctuations in the reservoirs, but does not account for the mean-field effect.

The mean-field effect of the Fe-tip manifests as an {\it effective} alternating magnetic field along the $\hat{\boldsymbol z}$ direction,
\begin{align}
B_{\rm tip}^{\rm eff}(t) &= \frac{1}{g \mu_B} 
\sum_k \big[ J_{ttkk\uparrow\uparrow}(t) - J_{ttkk\downarrow\downarrow}(t) \big]  f_{\beta}(\epsilon_{tk}) \nonumber \\ 
&= \frac{\chi \Gamma_t}{\pi g \mu_B} \int  \frac{D_t(\omega) }
{\omega  - \epsilon} f_\beta \big(\omega + V_{\rm dc} + V_{\rm ac}(t) \big) \,d\omega  
 \nonumber \\ 
&\simeq  B_0 + B_1 \sin(\omega_{\rm rf} t). \label{Beff}
\end{align}
Here, $\epsilon_{tk}$ refers to the electron energy in the $k$th band of the tip, and $f_\beta(\omega) = \frac{1}{e^{\beta (\omega - \mu)}+1}$ is the Fermi function with $\beta = 1/T$ denoting the inverse temperature. For brevity, we have set $\hbar = k_{\rm B} = e = 1$ and $\mu =0$. 
For the case studied in Fig.~\ref{fig2}(a), the magnitudes of $B_0$ and $B_1$ are on the order of $\sim 0.01\,{\rm T}$. 
It is important to emphasize that $B_{\rm tip}^{\rm eff}(t)$ does not arise from the modulation of electron hopping integrals, since the applied voltage does not alter their magnitudes \cite{SupMat1}. Instead, the alternating component of $B_{\rm tip}^{\rm eff}(t)$ originates from $V_{\rm ac}(t)$ through the modulation of the electronic band levels as well as the electrochemical potential of the tip. Details are given in the SM \cite{SupMat1}.

With the inclusion of the effective magnetic field, the Hamiltonian of the local spin is expressed as 
$H'_{\rm spin}(t) = H_{\rm spin} + \langle H_{\rm int}(t) \rangle_{\rm res} = H'_{\rm Zeeman} + H_{\rm Drive}(t)$. Here, $H'_{\rm Zeeman} = H_{\rm spin} + g \mu_B B_0 \hat{S}^z$ results in a renormalized Zeeman splitting, given by $\omega_L \simeq g \mu_B (|\boldsymbol B_{\rm ext}| + B_0 \cos\theta)$; and $H_{\rm Drive}(t) = g\mu_B B_{1} \hat{S}^z \sin(\omega_{\rm rf}t)$  serves as the driving source for the spin dynamics.

The spin dynamics is characterized by the temporal evolution of the reduced density matrix, $\rho(t) \equiv {\rm tr}_{\rm res}(\rho_{_{\rm T}})$. 
By utilizing the cumulant expansion for the total density matrix $\rho_{_{\rm T}}$
and the Born approximation $\rho_{_{\rm T}} \approx \rho_{\rm res}^{\rm eq} \otimes \rho(t)$ \cite{SupMat1}, we obtain a quantum master equation for $\rho(t)$: 
\begin{equation}
 \dot{\rho}(t) = -i \left[ H'_{\rm Zeeman} + H_{\rm Drive}(t), \rho(t) \right] + R(t). \label{eom}
\end{equation}
Here, the commutator describes the Larmor precession of the local spin, and $R(t)$ captures the influence of fluctuations in the reservoirs, leading to the dissipation and relaxation of the local spin.
In the following, we neglect $R(t)$ because of the weak magnitudes of $H'_{\rm int}(t)$ \cite{SupMat1}.
Equation~\eqref{eom} with $R(t)=0$ formally recovers the dynamical theory of the conventional ESR \cite{weil2007electron}.
This underscores the spin dynamics as the predominant source of the STM-ESR signal, thus addressing the first fundamental question raised in this Letter.

By performing the Laplace transform of Eq.~\eqref{eom}, 
we obtain the frequency-resolved reduced density matrix at the driving frequency $\omega_{\rm rf}$, $\tilde{\rho}(\omega+i\eta; \omega_{\rm rf}) = \int_0^\infty dt\, e^{i(\omega+i\eta) t} \rho(t)$ \cite{SupMat1}.
By setting $\omega =0$ and scanning the driving frequency $\omega_{\rm rf}$, $\tilde{\rho}(i\eta; \omega_{\rm rf})$ exhibits a resonance at $\omega_{\rm rf} = \omega_L$. The resonance is subject to power broadening \cite{schweiger2001principles}, and its linewidth varies linearly with the magnitude of $B_1$ \cite{SupMat1}. This is consistent with the experimental observation that the linewidths of STM-ESR spectra grow linearly with $V_{\rm rf}$ \cite{Bae2018Nov,PhysRevLett.122.227203,Seifert2020Oct}, a fact that the spin-transfer-torque mechanism \cite{PhysRevB.99.054434} fails to account for.

To address the second fundamental question on how the spin dynamics characterized by Eq.~\eqref{eom} gives rise to the spin resonance signatures in the electric current, we start with the expression for the total electric current, $I(t) = i\, {\rm tr}_{_{\rm T}} (H_{ts} \rho_{_{\rm T}}) + {\rm c.c.}$, where the trace is taken over the Hilbert space of the total system. Utilizing time-dependent perturbation theory \cite{mukamel1995principles,SupMat1}, we obtain the following expression for the time derivative of $I(t)$,
\begin{align}
\dot{I}(t)  & \simeq  i \, {\rm tr} \Big[  
  H_{\rm spin} \int_{0}^t d\tau \big\langle \overline{H_{ts}(t)} \,
\mathcal{U}(t,\tau) \, \overline{H_{st}(\tau)} \big\rangle_{\rm res} \, \rho(\tau)  \Big] \nonumber \\
 & \quad + {\rm c.c.} \label{dotI}
\end{align}
Here, ${\rm tr}$ denotes the trace over the Hilbert space of the local spin, and $\mathcal{U}(t,\tau) = e^{-i \overline{H_{\rm spin}} (t-\tau)} $ with $\overline{O}\,\ast \equiv [\hat{O}, \ast]$.

The STM-ESR spectrum is given by the dc component of the electric current, i.e., 
$\tilde{I}(\omega_{\rm rf}) = \lim_{t_f \to \infty} \frac{1}{t_f}\int_0^{t_f} dt\, I(t)$. 
By applying the Laplace transform to Eq.~\eqref{dotI} and setting $\omega =0$, we obtain a concise analytic formula, 
\begin{equation}
\tilde{I}(\omega_{\rm rf})  
= C \sin\theta\, \tilde{\boldsymbol S}(\omega_{\rm rf} ) \cdot {\boldsymbol e}_{\perp}. \label{stm-esr}
\end{equation}
Here, $\tilde{\boldsymbol{S}}(\omega_{\rm rf}) =  {\rm tr} \{\hat{\boldsymbol S}
\, \eta \tilde{\rho}(i\eta; \omega_{\rm rf}) \}_{\eta \to 0^+}$, 
and ${\boldsymbol e}_{\perp}$ is a unit vector perpendicular to ${\boldsymbol B}_{\rm ext}$ in the plane spanned by ${\boldsymbol B}_{\rm ext}$ and $\hat{\boldsymbol z}$.
In Eq.~\eqref{stm-esr}, the resonance signatures present in the spin dynamics $\tilde{\rho}(i\eta; \omega_{\rm rf})$ are directly passed on to the spectrum $\tilde{I}(\omega_{\rm rf})$. The prefactor $C$ originates from the correlation function 
$\big\langle \overline{H_{ts}(t)} \,\mathcal{U}(t,\tau) \, \overline{H_{st}(\tau)}  \big\rangle_{\rm res}$, and it represents the rate of electron transport across the junction through electron co-tunneling events:
\begin{equation}
C = \frac{\chi \Gamma_{s} \Gamma_{t}}{2\pi^2} \int \frac{D_t(\tilde{\omega}) D_s(\omega)}
{(\omega - \epsilon)^2} 
\left[ f_\beta(\tilde{\omega}) - f_\beta(\omega)\right] \,d\omega.  \label{prefactor}
\end{equation}
Here, $\tilde{\omega} = \omega + V_{\rm dc}$, indicating that the role of $V_{\rm dc}$ is to create an energy window for electron transport.

We now apply the analytic theory to decipher the essential features displayed in Fig.~\ref{fig2}. Firstly, Eqs.~\eqref{stm-esr} and \eqref{prefactor} reveal the conditions under which the STM-ESR signal vanishes: a spin-unpolarized tip ($\chi=0$), alignment of ${\boldsymbol B}_{\rm ext}$ with $\hat{\boldsymbol{z}}$ ($\theta=0$ or $180^\circ$), or when $V_{\rm dc} =0$. 
Secondly, both $\omega_L$ and $\tilde{I}(\omega_{\rm rf})$ are trigonometric functions of $\theta$. As evident in Fig.~\ref{fig2}(b), our analytic theory accurately reproduces the observed $\theta$-dependence of the peak position and amplitude in the numerical simulations.
Thirdly, as shown in Fig.~\ref{fig2}(c), the exponential decay of the signal with increasing temperature is accounted for by the statistical distribution encoded in $\tilde{\rho}$. Moreover, the nonlinear relationship between the signal and $V_{\rm rf}$ is explained by the amplification of $\tilde{\boldsymbol S}(\omega_{\rm rf})$ through $B_{\rm tip}^{\rm eff}$, along with the effects of fluctuations in the reservoirs.

The quantum master equation in Eq.~\eqref{eom} and the analytic formula in Eq.~\eqref{stm-esr} provide a comprehensive theoretical framework for describing the STM-ESR spectrum of an individual spin system.
Extending this framework to scenarios involving multiple adatoms or molecules is straightforward, which can be achieved by replacing the spin operator in Eq.~\eqref{stm-esr} with the operator corresponding to the magnetic center directly coupled to the tip.

{\it Conclusion.} The simulated STM-ESR spectra accurately and comprehensively reproduce the key features observed in experiments. This validates both the microscopic model representing the experimental setup and the understanding of the essential roles played by the bias voltage. Crucially, the presented analytic theory reveals that the STM-ESR signal originates from the net electron flow pumped by the Larmor precession of the local spin. Specifically, the effective magnetic field $B_{\rm tip}^{\rm eff}(t)$, which emerges as the mean-field influence of the spin-polarized tip, serves as the driving source for the spin dynamics. While spin dynamics are dominant, we do not exclude the possible existence of synergistic effects involving nuclear and charge dynamics \cite{baumann_electron_2015,PhysRevB.96.205420,Seifert2020Oct}. 
The theoretical insights presented in this Letter may be useful for designing innovative atomic-scale platforms for spin sensing and manipulation.

\begin{acknowledgments}

The support from the National Natural Science Foundation of China (Grant Nos. 22203083, 22393910 and 22425301), 
Innovation Program for Quantum Science and Technology (Grant No. 2021ZD0303305), 
and Strategic Priority Research Program of the Chinese Academy of Sciences (Grant No. XDB0450101) is gratefully acknowledged. 
The authors thank Xu~Ding and Jiaan~Cao for fruitful discussions. 

\end{acknowledgments}

\bibliographystyle{apsrev4-2}
%


%

%
\end{document}